\documentclass[journal=jpclcd,manuscript=article]{achemso}
\setkeys{acs}{maxauthors=0}

\usepackage{achemso}
\usepackage{graphics}
\usepackage{amssymb,amsfonts}
\usepackage{graphicx}
\usepackage[table]{xcolor}
\usepackage{caption}
\usepackage{subcaption}
\usepackage{colortbl}
\usepackage{amsmath}
\usepackage{amsopn}
\usepackage{siunitx}
\usepackage{bm}
\usepackage{color}
\usepackage{array}
\usepackage{lscape}
\usepackage{mciteplus}
\usepackage[version=3]{mhchem}
\usepackage{ulem}
\usepackage{listings}
\usepackage{enumerate}

\SectionNumbersOn

\author{Janus J. Eriksen}
\email{jeriksen@uni-mainz.de}
\affiliation[Johannes Gutenberg-Universit\"at Mainz]
{Institut f\"ur Physikalische Chemie, Johannes Gutenberg-Universit\"at Mainz, D-55128 Mainz, Germany}
\author{Filippo Lipparini}
\affiliation[Johannes Gutenberg-Universit\"at Mainz]
{Institut f\"ur Physikalische Chemie, Johannes Gutenberg-Universit\"at Mainz, D-55128 Mainz, Germany}
\altaffiliation{Present address: {\it{Dipartimento di Chimica e Chimica Industriale, Universit{\`a} di Pisa, I-56124 Pisa, Italy}}}
\author{J{\"u}rgen Gauss}
\affiliation[Johannes Gutenberg-Universit\"at Mainz]
{Institut f\"ur Physikalische Chemie, Johannes Gutenberg-Universit\"at Mainz, D-55128 Mainz, Germany}

\title[TITLE]{Virtual Orbital Many-Body Expansions: A Possible Route towards the Full Configuration Interaction Limit}

\begin{document}

%
%%%%%%%%%%%%%%%%%%%%%%%%%%%%%%%%%%%%%%%%%%%%%%%%%%%%%%%%%%%%%%%%%%%
%                                                                    				Abstract
%%%%%%%%%%%%%%%%%%%%%%%%%%%%%%%%%%%%%%%%%%%%%%%%%%%%%%%%%%%%%%%%%%%
%
\begin{abstract}

In the present letter, it is demonstrated how full configuration interaction (FCI) results in extended basis sets may be obtained to within sub-kJ/mol accuracy by decomposing the energy in terms of many-body expansions in the virtual orbitals of the molecular system at hand. This extension of the FCI application range lends itself to two unique features of the current approach, namely that the total energy calculation can be performed entirely within considerably reduced orbital subspaces and may be so by means of embarrassingly parallel programming. Facilitated by a rigorous and methodical screening protocol and further aided by expansion points different from the Hartree-Fock solution, all-electron numerical results are reported for H$_2$O in polarized core-valence basis sets ranging from double-$\zeta$ (10 $e$, 28 $o$) to quadruple-$\zeta$ (10 $e$, 144 $o$) quality.

\end{abstract}

\newpage

The full configuration interaction (FCI) wave function represents the exact solution to the electronic Schr{\"o}dinger equation within a given fixed-sized one-electron basis set. This formal attractiveness aside, its practical realization is generally impeded by a twofold curse of dimensionality~\cite{knowles_handy_fci_cpl_1984,olsen_fci_jcp_1988,olsen_fci_cpl_1990}; within a basis set of a certain quality, the scaling of the FCI model is exponential with respect to the number of electrons, and even for a fixed system size, the computational requirements grow exponentially with respect to the number of molecular orbitals (MOs). To circumvent this despairing intractability, various classes of approximations to the FCI model are usually invoked, not to mention the powerful approaches that derive from density matrix renormalization group (DMRG) theory~\cite{chan_dmrg_2011,wouters_dmrg_2014,knecht_dmrg_2016} and stochastic solutions to the Schr{\"o}dinger equation~\cite{booth_alavi_fciqmc_jcp_2009,booth_alavi_fciqmc_jcp_2010,booth_alavi_fciqmc_jcp_2011,booth_alavi_fciqmc_jcp_2017}. In the most popular and conventional of these classes, a truncation of the wave function expansion is enforced; this type of approach encompasses established and successful methods such as those of the configuration interaction (CI) and coupled cluster (CC) hierarchies~\cite{shavitt_bartlett_cc_book,mest}. Alternatively, one might conceive approximations aimed directly at the energy expression; in such approaches, the energy is initially decomposed followed by a feasible truncation. This change of target from the wave function to the energy motivates the use of many-body expansions (MBEs), which provide access to an incremental take on electron correlation phenomena. In recent years, computational strategies based on MBEs have experienced a notable rise in popularity~\cite{xantheas_mbe_int_energy_jcp_1994,paulus_stoll_phys_rev_b_2004,stoll_paulus_fulde_jcp_2005,truhlar_mbe_jctc_2007,truhlar_mbe_dipole_mom_pccp_2012,crawford_mbe_optical_rot_tca_2014,gordon_slipchenko_chem_rev_2012,parkhill_mbe_neural_network_jcp_2017,zgid_gseet_jpcl_2017}. However, whereas the objects entering these expansions have typically been the individual monomer molecules or molecular moieties of a supersystem, as, for instance, in the context of the local incremental scheme~\cite{stoll_cpl_1992,stoll_phys_rev_b_1992,stoll_jcp_1992,friedrich_jcp_2007}, these may also be chosen as the occupied spatial MOs of a system (labelled with indices $\{i,j,k,\ldots\}$), in which case the master equation becomes the so-called $N$th-order Bethe-Goldstone equation~\cite{nesbet_phys_rev_1967_1,nesbet_phys_rev_1967_2,nesbet_phys_rev_1968}
\begin{align}
E_{\text{FCI}} &= \sum_{i}\epsilon_{i} + \sum_{i>j}\Delta\epsilon_{ij} + \sum_{i>j>k}\Delta\epsilon_{ijk} + \ldots \nonumber\\
&= E^{(1)} + E^{(2)} + E^{(3)} + \ldots \label{bethe_goldstone_eq}
\end{align}
In the decomposition of the FCI energy, $E_{\text{FCI}}$, in Eq. \ref{bethe_goldstone_eq}, the changes in electron correlation (increments) from correlating the electrons of two orbitals over one ($\Delta \epsilon_{ij}$), three over two ($\Delta \epsilon_{ijk}$), etc., are given as
\begin{subequations}
\label{bg_increments_eqs}
\begin{align}
\Delta \epsilon_{ij} &= \epsilon_{ij} - (\epsilon_{i} + \epsilon_{j}) \label{two_body_inc} \\
\Delta \epsilon_{ijk} &= \epsilon_{ijk} - (\Delta \epsilon_{ij} + \Delta \epsilon_{ik} + \Delta \epsilon_{jk}) - (\epsilon_{i} + \epsilon_{j} + \epsilon_{k}) \ . \label{three_body_inc}
\end{align}
\end{subequations}
The calculation of order approximations to $E_{\text{FCI}}$ thus presupposes knowledge of the components of all contributions at lower orders, in the sense that lower-order increments enter the expressions for higher-order increments~\cite{harris_monkhorst_freeman_1992}. To $n$th order, $E^{(n)}$, or---in the present context---for the account of $2n$-electron correlation in the typical case of a closed-shell molecule, closed-form energy expressions exist in the literature~\cite{kaplan_mbe_mol_phys_1995,herbert_mbe_acc_chem_res_2014}, albeit only in the limit where the full orbital space remains untruncated ({\it{vide infra}}).

If the expansion in Eq. \ref{bethe_goldstone_eq} is left untouched, one does nothing but calculate the FCI energy in an immensely cumbersome fashion. However, and this was the main motivation behind Nesbet's earlier work in terms of generalized Bethe-Goldstone equations~\cite{nesbet_phys_rev_1967_1,nesbet_phys_rev_1967_2,nesbet_phys_rev_1968}, the expansion might become of practical value if contributions from higher-order combinations of orbitals (denoted as {\it{tuples}} in the present work) turn out to be negligible. In that case, the exact FCI limit may be approached---at least in principle---by correlating an increasing number of electrons independently and in succession. Here, it is worth noting that the two most celebrated features of CI, the orbital invariance and upper bound of the ground state energy, are in general sacrificed following any pragmatic truncation of Eq. \ref{bethe_goldstone_eq}. However, such a sacrifice will prove beneficial for the sake of being able to incrementally approximate $E_{\text{FCI}}$, {\it{if}} the total error with respect to a conventional result---which is anyways only obtainable in the most modest of basis sets---is sufficiently low. In the present work, the energetic tolerance, with which we will be concerned, is that of thermochemical (sub-kJ/mol) accuracy. Various schemes formulated around this fundamental idea have recently been proposed, such as the CCEMBE approach by Ruedenberg and Windus~\cite{bytautas_ruedenberg_ceeis_jcp_2004,bytautas_ruedenberg_ceeis_jpca_2010,ruedenberg_windus_mbe_jpca_2017} (albeit not targeted at the FCI limit) and notably the incremental FCI scheme by Zimmerman~\cite{zimmerman_ifci_jcp_2017_1,zimmerman_ifci_jcp_2017_2,zimmerman_ifci_jpca_2017}.

However, while $N$ may be small (as for, e.g., H$_2$O, in which case $N=10$), extended basis sets are compulsory for solving the Schr{\"o}dinger equation, and the FCI curse of dimensionality hence still prevails. Also, for code parallelization to be effective---an aspect that becomes increasingly important when developing novel algorithms that aim at embracing current as well as future supercomputer architectures---the total number of independent calculations must add up to a significant figure. Now, at any given order in Eq. \ref{bethe_goldstone_eq}, the number of individual calculations is determined by sheer combinatorics, and returning to the case of water, the total number of calculations to be distributed will thus be a fixed $\sum^{5}_{k=1}C(5,k) = 31$ (where $C(n,k)$ is a binomial coefficient), regardless of the choice of basis set. Furthermore, as the number of virtual MOs rises steeply along with an increase in basis set size, even low-order approximations to Eq. \ref{bethe_goldstone_eq} are soon to become unachievable. For this reason, we propose to turn things around by considering the objects of the MBE not to be the occupied, but rather the virtual MOs of the system. Thus, while a possible disadvantage of such an approach might be that some of the intuitive physical interpretation of the expansion itself is lost, clear advantages include the huge potential in terms of inherent massive parallelism as well as the fact that all basis sets become accessible for systems such as H$_2$O. Indeed, the number of independent calculations will now increase upon moving to larger basis sets, while the cost of the individual calculations remains marginal, operating under the assumption that Eq. \ref{bethe_goldstone_eq} still converges reasonably fast.

Having decided on virtual MOs as the expansion parameters in Eq. \ref{bethe_goldstone_eq}, the question remains as to whether such a procedure will in general be capable of eliminating the well-known redundancy of the FCI wave function~\cite{knowles_handy_fci_jcp_1989,ivanic_ruedenberg_ci_deadwood_tca_2001,bytautas_ruedenberg_ci_deadwood_cp_2009}. In general this is not so, and in order to avoid accounting for a colossal amount of vanishing contributions at various orders in the expansion, we have devised a rigorous screening protocol which is built into the expansion. In this way, the current algorithm strives towards being able to compress the set of variable parameters to the largest extent possible subject to an {\it{a priori}} threshold. Thus, the philosophy is akin to, but at the same time significantly different from that behind so-called selected and projector CI methods~\cite{malrieu_selected_ci_jcp_1973,harrison_selected_ci_jcp_1991,neese_selected_ci_talk_2016,stampfuss_wenzel_selected_ci_jcp_2005,head_gordon_whaley_selected_ci_jcp_2016,holmes_umrigar_heat_bath_ci_jctc_2016,sharma_umrigar_heat_bath_ci_jctc_2016,schriber_evangelista_selected_ci_jcp_2016,zhang_evangelista_projector_ci_jctc_2016}, and it may hence---on par with these---be viewed as a deterministic counterpart to stochastic FCI quantum Monte Carlo (FCIQMC)~\cite{booth_alavi_fciqmc_jcp_2009,booth_alavi_fciqmc_jcp_2010,booth_alavi_fciqmc_jcp_2011,booth_alavi_fciqmc_jcp_2017}.

%
%%%%%%%%%%%%%%%%%%%%%%%%%%%%%%%%%%%%%%%%%%%%%%%%%%%%%%%%%%%%%%%%%%%
%                                                                    		COMPUTATIONAL DETAILS
%%%%%%%%%%%%%%%%%%%%%%%%%%%%%%%%%%%%%%%%%%%%%%%%%%%%%%%%%%%%%%%%%%%
%

More specifically, the screening protocol proceeds in the following manner. At orders $k\leq3$, all possible complete active space CI (CAS-CI) calculations involving one, two, or three virtual and the complete set of occupied MOs are performed. At all subsequent orders, possible {\it{child}} tuples at order $k+1$ are generated from the complete set of {\it{parent}} tuples at order $k$ in a graph-like fashion. For each parent tuple at order $k$, denoted as $[a,b,\ldots,c]$, we probe whether or not to consider the child tuple at order $k+1$, $[a,b,\ldots,c,d]$, which is constructed by appending the parent tuple by an MO with index $d>c$. This is done by defining the following set of tuples of order (length) $k$
\begin{align}
\{\Lambda\}_{k} &= S_{k-1}\{[a,b,\ldots,c]\} \otimes \{[d]\} \label{lambda_set}
\end{align}
where the action of $S_{k-1}$ onto the parent tuple is to construct all possible subsets of length $k-1$, and the direct product produces all combinations that append the MO $d$ to any of these lists. The following condition now governs the potential screening of the child tuple, $[a,b,\ldots,c,d]$
\begin{align}
T_{k} < |\Delta\epsilon_{\lambda}| \hspace{0.25cm} \forall \hspace{0.25cm} \lambda \in \{\Lambda\}_{k} \label{screen_cond}
\end{align}
for some numerical energy threshold, $T_{k}$, see below. That is, if the orbital $d$ is sufficiently correlated with all combinations of orbitals present in the parent tuple, then said child tuple will be among the tuples that are considered at order $k+1$, and {\it{vice versa}}, if the condition in Eq. \ref{screen_cond} fails to be satisfied. The main assumption behind the screening protocol is thus that the increase in correlation from correlating the MOs of the parent tuple in the presence of the new MO will be minuscule to within the desired accuracy. Furthermore, the implications of the screening propagate implicitly to higher orders, as all potential child and grandchild tuples from $[a,b,\ldots,c,d]$ will automatically be neglected as well.

Now, while the graph-like generation of input tuples necessitates a tight threshold early on in the expansion, this is less decisive upon moving to higher orders if indeed the sum of the (increasingly manifold) individual energy increments becomes increasingly negligible. Thus, we might opt to relax the threshold along the expansion. Specifically, at order $k=1$, the threshold is fixed to a value of $T_1 = T_{\text{init}} \equiv \num{1.0e-10}$ a.u., which is the value to within which the energy of the individual CAS-CI calculations is converged, and hence a conservative lowest threshold for which the numerical precision of the calculation may be controlled~\cite{herbert_mbe_jcp_2014}. This is so, as all contributions with energy increments below this limit will ultimately be tainted from numerical noise. At all subsequent orders, however, the threshold takes the form
\begin{align}
T_{k} = T_{\text{init}} \cdot a^{k-1}
\end{align}
where $a \geq 1.0$ is a relaxation factor. For a fixed value of $T_{\text{init}}$, $a$ is the sole parameter defining our expansion. An important aspect when discussing any screening protocol, however, is concerned with the energy assembly at each order in the expansion, as the use of screening generally hinders the use of closed formulas~\cite{herbert_mbe_acc_chem_res_2014,herbert_mbe_jcp_2014} for summing up $E^{(n)}$. Instead, the direct recursive scheme in Eqs. \ref{bg_increments_eqs} is required for calculating the individual increments.

Finally, we note that we have the freedom to choose an arbitrary base for the expansion in Eq. \ref{bethe_goldstone_eq}, in particular one that is different from the Hartree-Fock (HF) solution. For instance, we may let the expansion target the gap in correlation energy between either the second-order M\o ller-Plesset (MP2)~\cite{mp2_phys_rev_1934} or CC singles and doubles (CCSD)~\cite{ccsd_paper_1_jcp_1982} solution and FCI instead of the full FCI correlation energy. While this assumes that an MP2 or CCSD calculation can be performed for the full system prior to the actual start of the expansion, as well as within each of the CAS spaces of the individual tuple calculations, the clear advantage of using such an intermediate model is that the individual energy increments are bound to be significantly smaller in value, leading to a potentially faster convergence towards the FCI solution. Furthermore, whenever an MP2/CCSD energy calculation is possible for the full system, one may additionally also diagonalize the virtual-virtual block of the 1-particle density matrix at that level of theory to obtain a set of virtual natural orbitals (NOs), which in turn allows for a more effective screening over the use of standard canonical virtual HF orbitals~\cite{lowdin_nat_orb_fci_phys_rev_1955,lowdin_nat_orb_fci_phys_rev_1956}. In the following, our choice of base model (HF, MP2, or CCSD) will implicitly also dictate the choice of virtual MO representation in the expansion (canonical orbitals or MP2/CCSD NOs, respectively). For the occupied MOs, on the contrary, any rotation of these among each other is redundant, and we will hence make use of canonical occupied HF orbitals throughout for all of the reported calculations.

In the present work, all MBE-FCI calculations have been performed using a novel code written exclusively in Python/NumPy~\cite{numpy}, of which all program phases have been explicitly parallelized using the message passing interface (MPI) protocol via its implementation in the {\sc{mpi4py}} Python module~\cite{mpi4py_1,mpi4py_2,mpi4py_3}. This extension hence allows for all computational tasks to be distributed in a parallel manner among a group of processes on a large computer cluster. The individual CAS-CI calculations have been performed using the Python-based {\textsc{pyscf}} platform~\cite{pyscf}\footnote{Git hash for the version of {\textsc{pyscf}} used herein: {\texttt{4a07bb9e0a}}}, with initial testing and verification enabled through an interface to the {\sc{cfour}} quantum chemical program package as backend engine~\cite{cfour}.

%
%%%%%%%%%%%%%%%%%%%%%%%%%%%%%%%%%%%%%%%%%%%%%%%%%%%%%%%%%%%%%%%%%%%
%                                                                    				RESULTS
%%%%%%%%%%%%%%%%%%%%%%%%%%%%%%%%%%%%%%%%%%%%%%%%%%%%%%%%%%%%%%%%%%%
%

%
\begin{figure}[ht]
        \centering
                \includegraphics[scale=0.85]{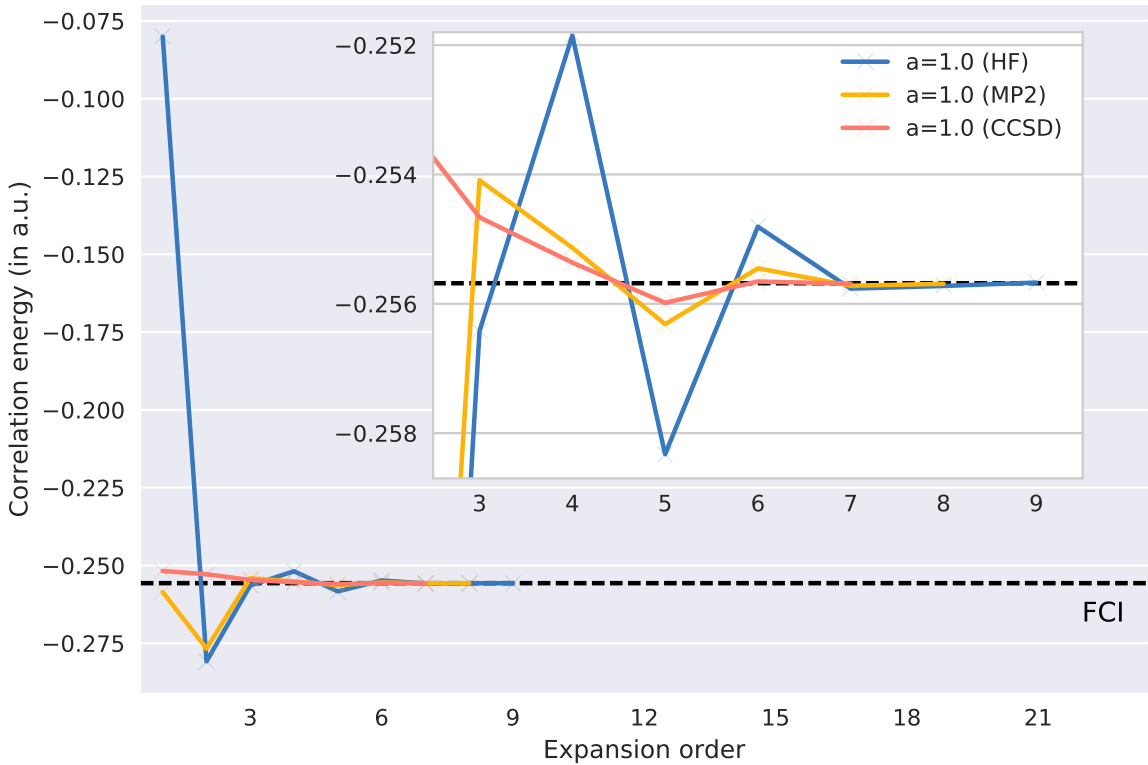}
        \caption{All-electron MBE-FCI/cc-pCVDZ results for H$_2$O with HF, MP2, and CCSD as the base for the expansion. The reference FCI result is indicated by the dashed line in black color.}
        \label{abs_energy_dz_fig}
\end{figure}
In Figure \ref{abs_energy_dz_fig}, all-electron MBE-FCI results are presented for H$_2$O ($R = 0.957$ \AA, $\angle = 104.2^{\circ}$) in a cc-pCVDZ core-valence basis set~\cite{dunning_1_orig,dunning_5_core} (10 $e$, 28 $o$). In comparison, a conventional calculation, making full use of Abelian point group symmetry ($C_{2\text{v}}$), would involve in excess of two billion variational parameters, which is close to the computational limit within the scope of anything but the most modern FCI implementations~\cite{nwchem_mcscf_impl_arxiv_2017}. As may be recognized by comparing the convergence of the three different curves in Figure \ref{abs_energy_dz_fig}, the use of an intermediate base model significantly improves the convergence rate towards the conventional FCI result. The MP2 model, which perturbatively accounts for all double excitations, already reduces the oscillations present in the HF-based curve, and the same observation---even more pronouncedly---is also true in moving from an MP2 to a CC expansion base. Furthermore, the use of a dynamic threshold ($a > 1.0$) is observed not to have any influence on the final result, as may be seen from the detailed comparison of the correlation energies (to within 5 decimal points) in Table \ref{results_table}, which also features FCI (as calculated using the CAS-SCF module in {\textsc{cfour}}~\cite{lipparini_gauss_rel_casscf_jctc_2016}) and high-level CC (CCSDT and CCSDTQ, as calculated via the interface to {\textsc{mrcc}} in {\textsc{cfour}}~\cite{mrcc,kallay_string_based_cc_jcp_2001}) reference data where available.

\begin{figure}[ht]
        \centering
                \includegraphics[scale=0.85]{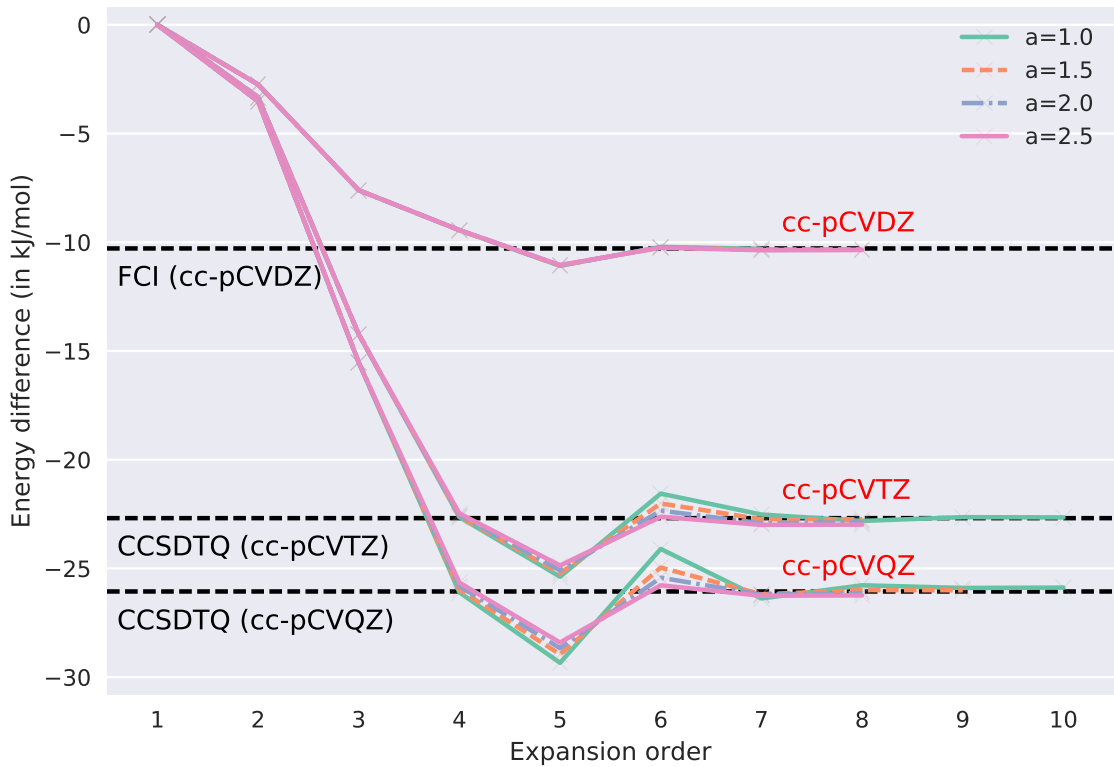}
        \caption{Recovery of the CCSD--FCI energy difference for H$_2$O in the cc-pCV$X$Z basis sets. Reference FCI and CCSDTQ results are indicated by dashed lines in black color.}
        \label{collected_fig}
\end{figure}

\begin{table}[ht]
\begin{center}
\begin{tabular}{l|r|r|r}
\hline\hline
\multicolumn{1}{c|}{Expansion} & \multicolumn{3}{c}{Basis set} \\
\cline{2-4}
\multicolumn{1}{c|}{threshold} & \multicolumn{1}{c|}{cc-pCVDZ} & \multicolumn{1}{c|}{cc-pCVTZ} & \multicolumn{1}{c}{cc-pCVQZ} \\
\hline\hline
$a = 1.0$ & $-0.25569$ & $-0.33283$ & $-0.35636$ \\
$a = 1.5$ & $-0.25570$ & $-0.33286$ & $-0.35641$ \\
$a = 2.0$ & $-0.25570$ & $-0.33291$ & $-0.35646$ \\
$a = 2.5$ & $-0.25571$ & $-0.33295$ & $-0.35650$ \\
\hline
CCSDT & $-0.25520$ & $-0.33250$ & $-0.35603$ \\
\hline
CCSDTQ & $-0.25566$ & $-0.33284$ & $-0.35643$ \\
\hline
FCI & $-0.25568$ & \multicolumn{1}{c|}{N/A} & \multicolumn{1}{c}{N/A} \\
\hline
$K_{\text{det}}$ & $\sim \num{2e09}$ & $\sim \num{4e13}$ & $\sim \num{5e16}$ \\
\hline
\hline
\end{tabular}
\end{center}
\caption{Total CCSD-based MBE-FCI/cc-pCV$X$Z correlation energies (in a.u.) for H$_2$O, as converged to within an uncertainty of $0.1$ kJ/mol ($\num{3.8e-5}$ a.u.). In addition, the number of determinants, $K_{\text{det}}$, entering a conventional FCI calculation as well as reference CC and/or FCI results are presented for comparison.}
\label{results_table}
\end{table}
Next, we turn to the considerably larger calculations within the cc-pCVTZ (10 $e$, 71 $o$) and cc-pCVQZ (10 $e$, 144 $o$) basis sets. In Table \ref{results_table}, MBE-FCI results are again presented for static as well as dynamic expansion thresholds. As was the case for the cc-pCVDZ calculations above, the use of threshold relaxation is observed only to affect the overall accuracy of the present scheme marginally, in comparison to the sub-kJ/mol precision at which we are aiming. Also, the overall convergence pattern remains relatively unchanged in the transition to larger basis sets, which is clear from Figure \ref{collected_fig}, in which the energetic difference between the CCSD base model and FCI is depicted {\it{vis-{\`a}-vis}} for all three basis sets. This is in perfect accordance with chemical intuition, in the sense that correlation as a whole is inherently a system- rather than a basis set-specific phenomenon.

\begin{figure}[ht]
        \centering
                \includegraphics[scale=0.85]{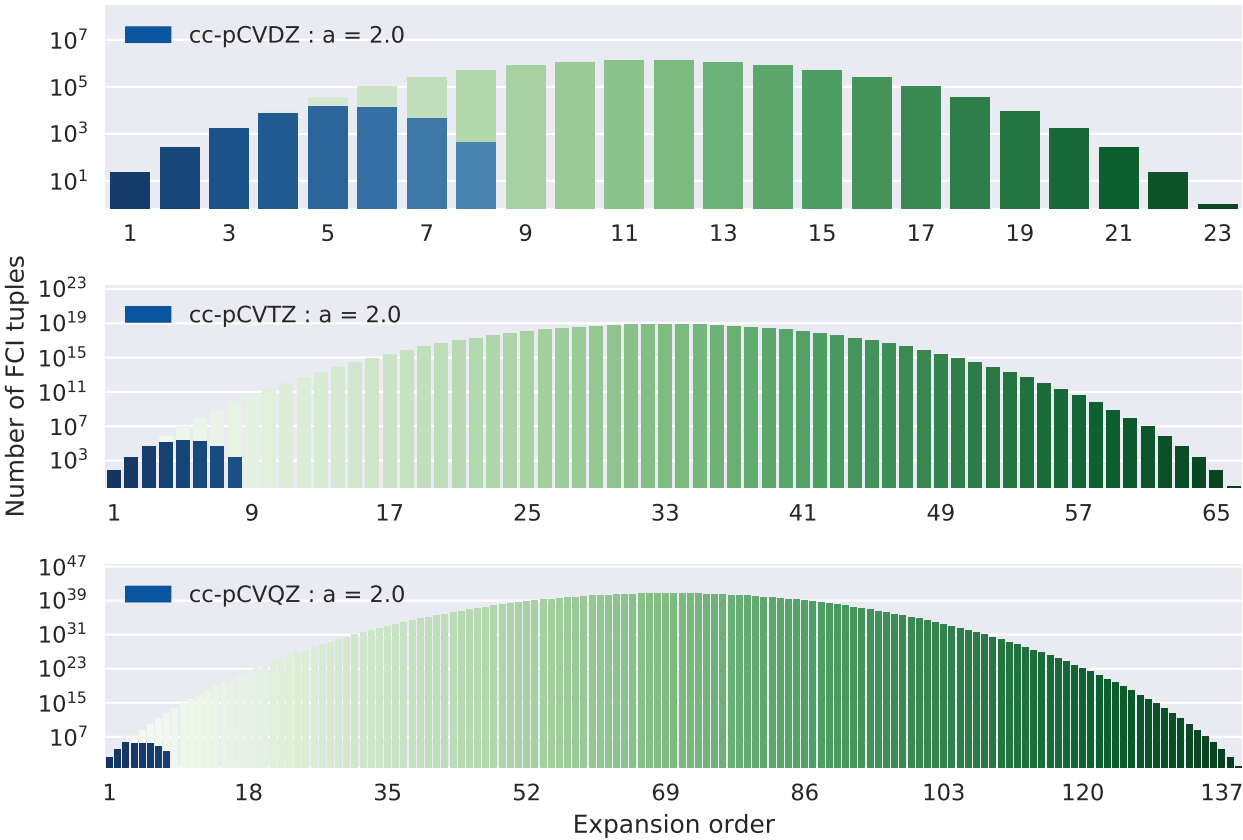}
        \caption{Comparison of the calculated number of tuples at each order (blue color) against the theoretical value (green color) for H$_2$O in each of the cc-pCV$X$Z basis sets.}
        \label{savings_fig}
\end{figure}

Finally, we briefly comment on the computational cost associated with the current algorithm. Allowing for threshold relaxation trivially results in a reduction of the total time-to-solution, as (i) fewer calculations need to be performed at each order and (ii) even fewer, if any, calculations need to be performed at high orders in the expansion. In Figure \ref{savings_fig}, we depict the number of orbital tuples that need be evaluated at each order in the expansions using a threshold relaxation of $a = 2.0$. As is clear from the comparison to the theoretical number of calculations, the savings with respect to a conventional calculation grow dramatically with increase in basis set size, as is particularly manifest in light of the fact that conventional FCI results are hypothetical for the cc-pCVTZ and cc-pCVQZ basis sets, due to the sheer size of the variational space (cf. Table \ref{results_table}). In terms of the accumulated number of tuples, the calculations in Figure \ref{savings_fig} involved a total of $41$k (cc-pCVDZ), $582$k (cc-pCVTZ), and $1302$k (cc-pCVQZ) individual calculations, and the relative increase in required tuples is hence observed not to increase proportionally to basis set size, but rather appear to saturate for higher cardinal numbers. Using our pilot implementation, the calculations required (in hours:minutes format) 00:13, 02:47, and 33:40 of walltime, respectively, on two nodes with 28 cores @ 2.4 GHz and 256 GB of memory each. In the transition from a cc-pCVDZ to a cc-pCVTZ basis set, the time ratio between the two calculations ($13.3$, using exact timings) corresponds satisfactorily well with the relative increase in individual tuples ($14.0$, using exact number of tuples). Moving to the even larger cc-pCVQZ basis, however, the relative increase in time is significantly worse. For this increase in time to solely reflect the corresponding increase in individual calculations, a communication bottleneck related to the handling of CAS space two-electron integrals for large basis sets remains to be resolved. The necessary modifications to the code required for resolving this issue are currently being implemented.

%
%%%%%%%%%%%%%%%%%%%%%%%%%%%%%%%%%%%%%%%%%%%%%%%%%%%%%%%%%%%%%%%%%%%
%                                                                    		CONCLUSIONS AND OUTLOOK
%%%%%%%%%%%%%%%%%%%%%%%%%%%%%%%%%%%%%%%%%%%%%%%%%%%%%%%%%%%%%%%%%%%
%

%
In the present letter, we have revisited the application of the MBE to the calculation of FCI energies, known as the so-called $N$th-order Bethe-Goldstone equation. By considering the objects of the equation not as the occupied MOs of the system at hand, but rather the virtual MOs, we have been able to extend the application range of FCI, while maintaining thermochemical (sub-kJ/mol) accuracy in comparison with the exact result. These enhancements have been made possible through the development of a simple, yet methodical screening protocol as well as the use of expansion points different from the HF solution.
Enabled by an all-Python/NumPy implementation of the new algorithm, we have presented all-electron results for H$_2$O in polarized core-valence basis sets ranging from double-$\zeta$ (10 $e$, 28 $o$) to quadruple-$\zeta$ (10 $e$, 144 $o$) quality.

However, we remark here that an MBE-based approach to the FCI electron correlation problem, in its current incarnation, will introduce a bias towards single-determinant dominated systems such as H$_2$O. To alleviate this hindrance of the general procedure, one may take advantage of the fact that the MBE allows for other choices of underlying references than the generic restricted HF solution. For instance, the ability to use open-shell HF references is work in progress within the existing computational framework. Alternatively, and this is also a current research field, one may extend the concept of MBEs even further by devising so-called dual (combined) expansions, in which MBEs are performed in both the occupied and the virtual MO space. In particular, one may perform an MBE in the set of occupied MOs, and then for each single orbital and orbital pair, triple, etc., generate a specific set of correlating virtual NOs. Such an approach will be capable of eliminating the factorial scaling with the number of electrons, which still restrains the current algorithm, under the assumption that occupied MBEs generally converge rapidly.

%
%%%%%%%%%%%%%%%%%%%%%%%%%%%%%%%%%%%%%%%%%%%%%%%%%%%%%%%%%%%%%%%%%%%
%                                                                    			     ACKNOWLEDGMENT
%%%%%%%%%%%%%%%%%%%%%%%%%%%%%%%%%%%%%%%%%%%%%%%%%%%%%%%%%%%%%%%%%%%
%
\section*{Acknowledgments}

The authors wish to thank assistant professor Dr. Lan Cheng of Johns Hopkins University for fruitful discussions and for supplying the CCSDTQ/cc-pCVQZ result for H$_2$O reported herein. Furthermore, J. J. E. wishes to thank Dr. Qiming Sun of California Institute of Technology for helpful discussions on the inner workings of the {\textsc{pyscf}} program. J. J. E. and F. L. are both grateful to the Alexander von Humboldt foundation for financial support.

\newpage

\providecommand*\mcitethebibliography{\thebibliography}
\csname @ifundefined\endcsname{endmcitethebibliography}
  {\let\endmcitethebibliography\endthebibliography}{}

\end{document}